# Evaluation of different WRF microphysics schemes: severe rainfall over Egypt case study

Muhammed ElTahan and Mohammed Magooda

Aerospace Engineering, Cairo University, Cairo, Egypt

Corresponding Author: muhammed_eltahan@hotmail.com

## Abstract

Egypt faced heavy rainfall starting from 26$^{th}$ October 2016 which caused flooding in Upper Egypt and Red Sea Coast especially the city of Ras Gharib. In this work, Weather Research and Forecasting (WRF) model was used with one domain of 10-km resolution for three days starting from 26$^{th}$ October 2016 to simulate this severe rainy event in terms of precipitation, temperature, pressure and wind speed. In order to use the WRF model several schemes should be configured, one of them is the microphysics scheme. Microphysics is the process by which moisture is removed from the air, based on other thermodynamic and kinematic fields represented within the numerical models.

Sensitivity of ten microphysics schemes were tested (Kessler, Lin, WRF single moment 3 and 5 class, Eta (Ferrier), WRF single Scheme 6 Class, Goddard, Thompson, Milbrandt-Yau Double Moment and Morrison double moment). The output temperature and precipitation from the model were compared to the satellite data obtained from Moderate Resolution Imaging Spectroradiometer (MODIS) and Tropical Rainfall Measuring Mission (TRMM), respectively. Our analysis showed that the model has the capability to simulate the rainy event spatially and produce the cloud pattern. Three coastal areas were chosen for this study, the city of Ras Gharib, Wadi Elgemal National park, and Elba National park. The results showed that the model generally tends to overestimate the precipitation for all the ten schemes and underestimated temperature compared to TRMM and MODIS data respectively.

**Keywords:** Precipitation, WRF, Microphysics, MODIS, TRMM

# 1- Introduction

Extreme weather events have negative impacts on cities, roads, communications, and air traffic which consequently leads to catastrophic effects on the different aspects of people's lives, and economy, Figure (1). By nature, more than 90 % of Egypt's land is desert, nonetheless Egypt is susceptible to multiple severe rainfall events over its coastlines every year with effects extending sometimes over the entire country. The need for monitoring and prediction system increased to help the authorities take preventive measures. Different numerical weather models alongside observations from both ground and satellite data are the corner base for such systems. Weather Research Forecast (WRF) model shows the capability to simulate heavy rainy events; so that it can be considered as component of alarm warning system (M.S. El-Sammany, 2010). Regional Climate Model (RegCM3) shows its capabilities for simulating both the spatial distribution and the amount of the rainfall over Egypt after validating the model output verse the observations (A.Esawy, 2011). Another severe event invaded Sinai Peninsula in Egypt during January 18, 2010 with very heavy rain and WEF was also to able to simulate this severe event (El Afandi, et al 2013).The reliability on WRF model for rainfall forecast has been increased (S. Ibrahim and El Afandi 2014). A research work started to catch the statistical characteristics of rainfall events in Egypt based on historical maximum daily rainfall records for 30 stations throughout the Egypt was provided by (Tamer A. Gado, 2017).

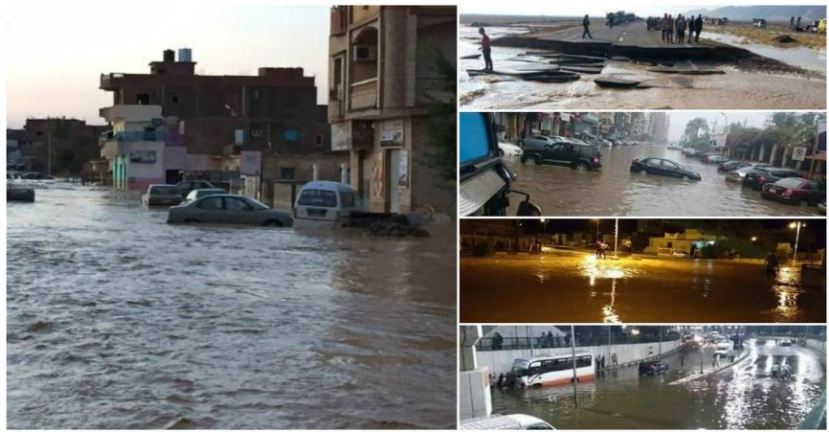

Figure 1: Damage from severe rain fall in 26$^{th}$ October 2016

Microphysics schemes in numerical models play important role in simulating formation of cloud droplets and fallout as precipitation and Land surface temperature (LST). It is also considered a key parameter in many fields hydrological, meteorological and environmental studies because it merges the interactions between the surface - atmosphere and energy fluxes between the atmosphere and the surface (Anderson, Norman, Diak, Kustas, & Mecikalski, 1997; Anderson et al., 2011; Wan & Dozier, 1996; Weng, Lu, & Schubring, 2004; Zhou et al., 2003). In this paper testing the effect of microphysics scheme on simulating severe rainfall event over Egypt in 26$^{th}$ October 2016 that extends for two days was investigated and model performance was validated against satellite observations obtained from both Tropical Rainfall Measuring Mission (TRMM) and Moderate Resolution Imaging Spectroradiometer (MODIS) over three coastal regions.

The paper is divided into 5 sections. Section one is the introduction. In Section two describes the modeling framework. The third section presents the satellite observation used in the validation process. The fourth section explains the results and Discussions. The last section states the conclusions and future work.

## 2- Modeling System Framework

Weather Research Forecasting (WRF) framework is the used one in this study. It is considered one of the common tool used in both research and operational applications. It is an efficient framework for atmospheric simulation system in different scales . WRF is as a common tool for both research and operational communities (NCAR technical note,2008) . The WRF was developed across different research centers check them at (NCAR technical note ,2008). The WRF Software Framework (WSF) works as an incubator for the dynamics solvers, physics packages that deals directly with the solvers, programs for initialization, chemistry and WRF Data Assimilation module (WRFDA) . There are two dynamics solvers (Both of them are Eulerian mass dynamical cores with terrain –following vertical coordinates) in the WSF (a) The Advanced Research WRF (ARW) solver development and Support primarily at NCAR / MMM (b) The NMM (Non hydrostatic Mesoscale Model) solver developed at NCEP/ EMC. Community support is provided by NCAR/DTC. (NCAR Technical note ,2008) The main Components of WRF used in this study are the WRF Preprocessing System (WPS), and ARW solver.

**The WRF Preprocessing System** (**WPS**) is one of the most important components of WRF. WPS can design the domain experiment, prepare geographical information WRF needs, convert meteorological data into a format WRF can use and prepare data to be used by WRF.

WPS is composed of three main programs whose collective role is to prepare input to the real program for real-data simulations. Each of the programs performs one stage of the preparation (a) geogrid program: related to processing geographical information based on the required domain for Analysis (b) ungrib program: this program converts meteorological data from GRIB format into the so-called "WRF Intermediate Format" (WIF). WIF is a format that the next program, metgrid.exe, understands and able to use (c) metgrid program: this program processes the WIF data and interpolates them horizontally. It then prepares the data to be used by WRF (User guide ARW -V3.8). The WRF-ARW core is based on an Eulerian solver for the fully compressible non-hydrostatic equations, cast in flux (conservative) form, using a mass (hydrostatic pressure) vertical coordinate. This solver use mass of dry air , Velocities , temperature and geopotential as prognostic variables while the non conservative variables such pressure could be derived from the prognostic variables. The solver also uses different numerical schemes such a 2nd- order time integration scheme coupled with third-order Runge-Kutta time-integration scheme for both gravity wave and acoustic modes . (Klemp,J.B. ,2007, Skamarock ,W.C. 2007, Skamarock, W. C., 2006, Skamarock, W. C., 2004, Wicker, L. J,2002 )

### 2-1 Model Setup

This study uses version 3.8 of WRF to simulate meteorology over model domain with terrain shown in Figure (2). The model domain is defined as Lambert Projection extends from 24 E to 36 E (120 grid points) in the east - west direction and from about 21.5 N to 32.5 N (120 grid points) in the north - south direction at horizontal grid $10 \times 10$ Km$^2$. The vertical grid consists of 41 level from surface up to 10 hpa. The static geographical fields, such as terrain height, soil properties, vegetation fraction, land use and albedo, etc., are interpolated and prepared to the model domain by using geogrid program in the WRF preprocessing system (WPS).

The initial and lateral boundary conditions for the meteorological fields are obtained from the National Center for Environmental Predictions (NCEP). Final Analysis (FNL) fields available every 6 hours at a spatial resolution of $1° \times 1°$. They were interpolated to the model domain by using ungrib and metgrid programs in the WRF preprocessing system (WPS).

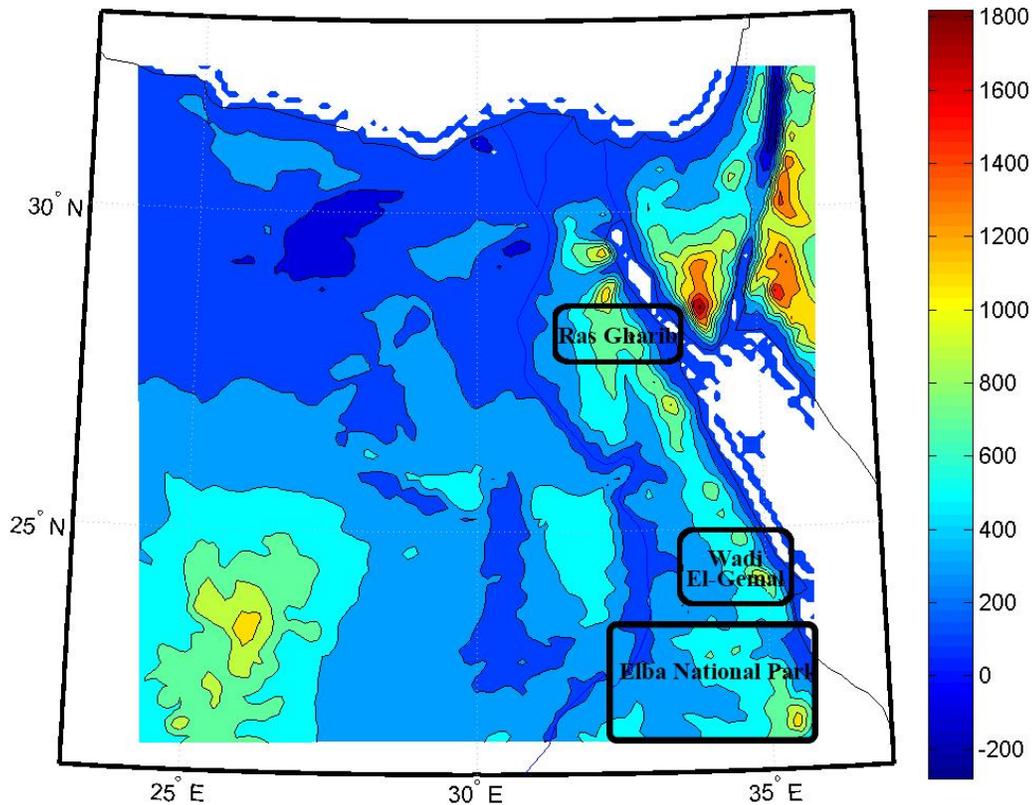

Figure 2: WRF Terrain map of Egypt showing
the locations of the three selected domains for validation

**2-3 Physical Parameterizations:**

The Following Physical schemes are used to configure the model and summarized in table (1). Microphysics behavior are selected interchangeably between the ten microphysics schemes Kessler, Lin, WRF single moment 3 and 5 class, Eta (Ferrier), WRF single Scheme 6 Class, Goddard, Thompson, Milbrandt-Yau Double Moment and Morrison 2 moment (E. Kessler, 1995;Y.L. Lin et al, 1983 ; S.Y.Hong et al ,2004 ; S.Y.Hong et al ,2006 ; W.K. Tao, 1989; G . Thompson. 2014; J. A. MILBRANDT and M. K. YAU ,2005(a) (b); H. MORRISON, 2009)

The Rapid Radiative Transfer Model (RRTMG) for shortwave and long-wave radiation is selected for the aerosol direct radiative effect (Mlawer et al., 1997). And since the Land Surface interact strongly with the atmosphere at all the scales. The Unified Noah Land surface had been selected to represent the land surface interaction (M. Tewari, 2004). The MM5 Similarity Scheme was used to describe the surface layer physics (Paulson, C. A 1970). Yonsei University Scheme (YSU) is chosen to represent the Planetary Boundary Layer (Hong et al, 2004) . The Cumulus Parameterization scheme is Grell 3D Ensemble Scheme (Georg A. Grell, 2002) .

| Scheme | Option number | Model |
|---|---|---|
| Microphysics | [1,2, 3, 4, 5, 6, 7, 8, 9,10] | Kessler<br>Lin (Purdue)<br>WRF single moment 3 (WSM3)<br>WRF single moment 5 (WSM5)<br>Eta (Ferrier)<br>WRF single Scheme 6b (WSM6)<br>Goddard<br>Thompson<br>Milbrandt-Yau Double Moment<br>Morrison Double Moment |
| Long-wave radiation | 4 | RRTMG |
| short wave radiation | 4 | RRTMG |
| Land surface | 2 | Unified Noah Land surface |
| PBL model | 1 | Yonsei University |
| Surface Layer | 1 | MM5 similarity |
| Cumulus | 5 | Grell 3D Ensemble Scheme |

Table (1) summary of model configuration

## 2-4 Microphysics behavior

Microphysics processes control the simulation of the formation of cloud droplets , the formation of ice crystals, their growth, and their fallout as precipitation. The key role for it on cloud, climate and weather models can be shown in (1) Latent heating/cooling: condensation, evaporation, deposition, sublimation, freezing, melting (2) Condensate loading: (mass of the condensate carried by the flow (3) Precipitation: fallout of larger particles. (4) Coupling with surface processes: moist downdrafts leading to surface-wind gustiness, cloud shading. (5) Radiative transfer: mostly mass for absorption/emission of long wave, particle size also important for Shot wave radiation. (6) Cloud-aerosol-precipitation interactions: aerosol affect clouds: indirect aerosol effects, but clouds process aerosols.. Microphysics schemes can be broadly categorized into two types (1) BULK: the Size distribution assumed to follow functional form (1-a) Inverse exponential (Marshall and Palmer,1948).(1-b) Gamma distribution. The advantage if this type is fewer number of prognostic variables equal which equal to cheap computational. (2) Bin: Size distribution discretizied into bins. All the available tested microphyucs schemes in WRF are bulk. Many different approaches have been used to examine the impact of microphysics on precipitation processes associated with convective systems. For example, ice phase schemes were developed in the 80's (Lin et al, 1983; Cotton et al. 1982, 1986; Rutledge and Hobbs 1984), and the impact of those ice processes on precipitation processes associated with deep convection were investigated (Yoshizaki 1986; Nicholls 1987; Fovell and Ogura 1988; Tao and Simpson 1989).

## 3-Satellite Observations

The Validation of the model's performance for predicting the parameters of this severe event was based on both MODIS and TRMM data for temperature and precipitation validation respectively.

## 3-1 MODIS

Santa Barbara Remote Sensing built the scientific instrument moderate-resolution imaging spectroradiometer (MODIS) . and lunched it to space by National Aeronautics and Space Administration (NASA) on both Terra and Aqua satellites in 1999 and 2002 respectively. This instrument has high capabilities to capture the data in terms of spectral and spatial resolution. The spectral and spatial resolutions are up to 36 band form 0.4 μm to 14.4 μm and 2 bands at 250 m, 5 bands at 500 m and 29 bands at 1 km respectively. Both MODIS Instruments on Terra and Aqua could cover the earth in 1 to 2 days . Both land and atmospheric Temperature could be extracted from bands 20 to 25 with spatial resolution 1 km. The Product land surface temperature (LST) Version 6 gives accurate estimates of LST at the bare soil or sand sites according to the evaluation of LST Algorithm for version 6 (Wan, 2014)MODIS/Aqua Land Surface Temperature and Emissivity level 2 ( MYD11_L2) product collection 6 has been used in this study ( Z. Wan, S. H. 2015).

## 3-2    TRMM

To study rainfall for weather and climate research, NASA and the Japan Aerospace Exploration (JAXA) launched the joint mission Tropical Rainfall Measuring Mission (TRMM) . From 1997 till 2015 the satellite program TRMM was able to estimate the global tropical rainfall (Kummerow *et al*., 2000). Five instruments were carried on TRMM. Two sensors related to cloud and radiation Lightning Imaging Sensor (LIS)  and The Clouds and the Earth's Radiant Energy System (CERES). Another three sensors related to rainfall  Precipitation Radar (PR) Microwave Imager (MI) and The Visible and Infrared Scanner  (VIRS).The TRMM dataset became the space standard for measuring precipitation. To verify the TRMM output products the Global Precipitation Measurement(GPM) Mission's Core Observatory was launched in February 2014  in addition to several ground validation programs were developed to make a comparison with rain gauges, ground radars. TRMM can provides sufficient estimate for rainfall although that its accuracy depends on retrieval algorithm (Gu *et al*., 2010).  The Daily precipitation data derived from 3B42_daily has been used in this study .

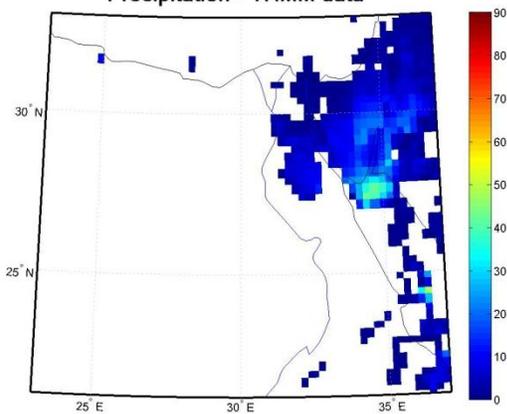

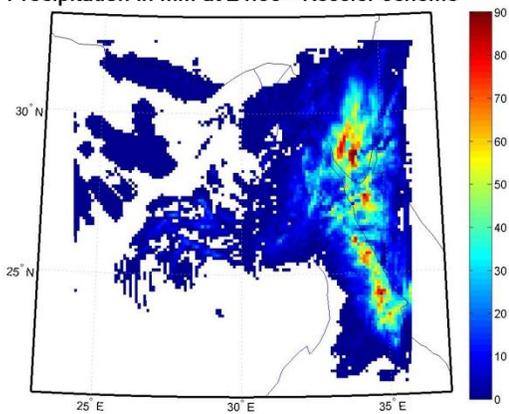
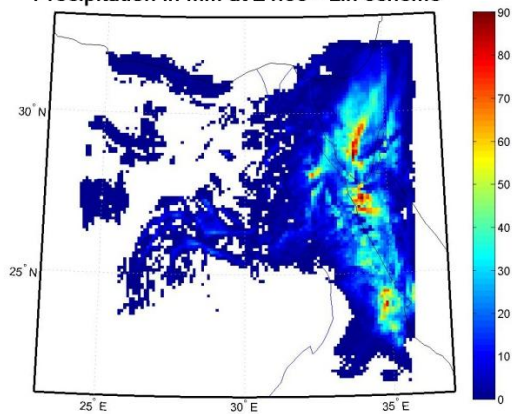

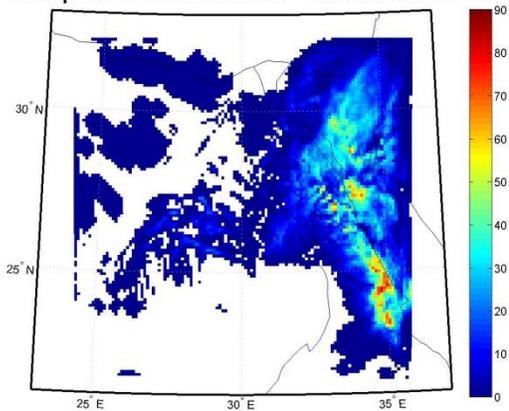
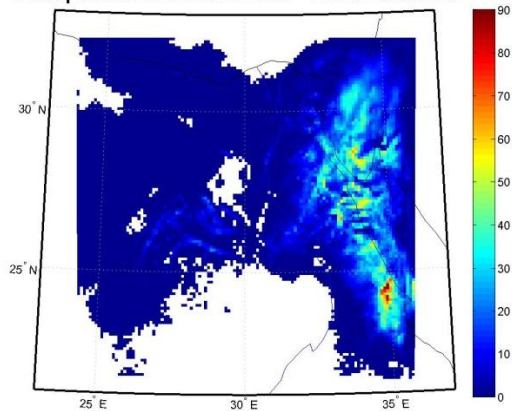

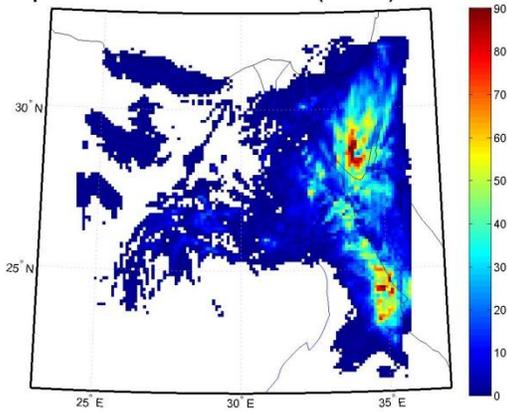
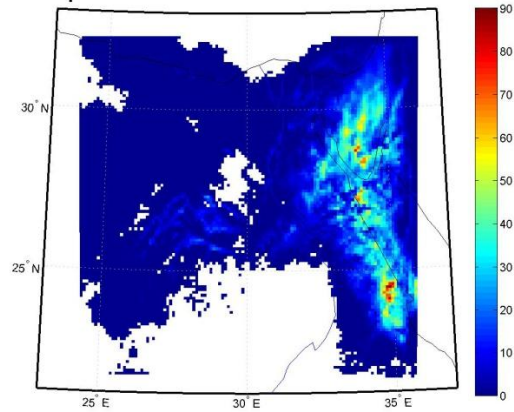
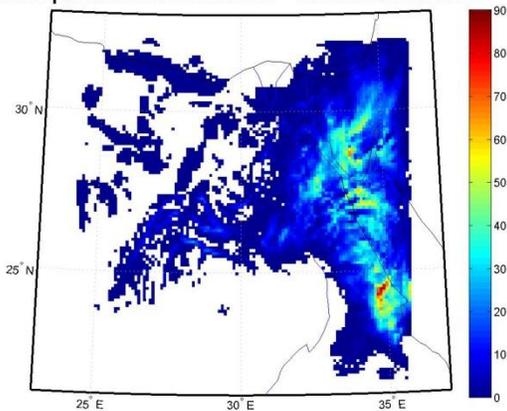
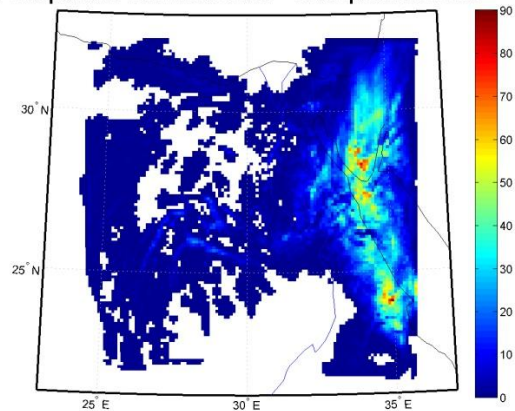
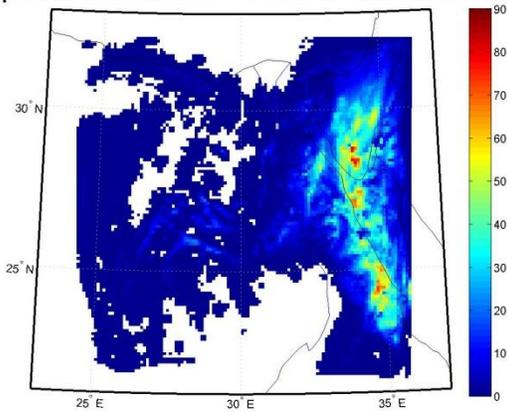
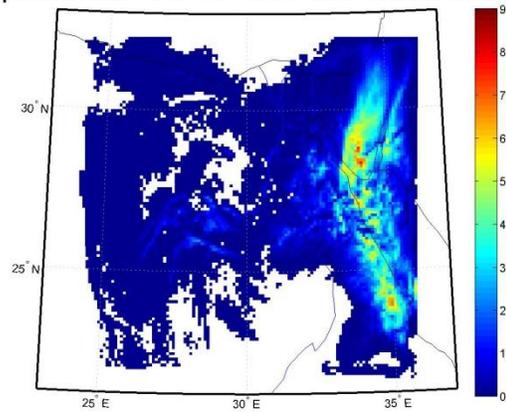

Figure (3): Spatial distribution for precipitation from TRMM satellite data and ten microphysics (Kessler, Lin ,WSM3, WSM5, Eta, WSM6, Goddard, Thompson. Milbrandt-Yau Double Moment and Morrison Double Moment Schemes)

## 4- Results and Discussions

Three selected domains with square boundaries over Ras Gharib, Wadi Elgemal National park and Elba National park to evaluate model performance are shown in table (2) and figure (2):

|  | Upper boundary | Lower boundary | Left boundary | Right boundary |
|---|---|---|---|---|
| Ras Gharib | 29 | 27.5 | 32 | 34 |
| Wadi Elgemal National park | 24.83 | 24.1 | 33.9 | 36.46 |
| Elba National park | 23.9 | 21.97 | 33.9 | 37.6 |

Table (2) Selected domains with boundaries

The spatial distribution for precipitation from ten different microphysics and TRMM data are shown in Fig (3) at midnight. Generally, all the schemes successfully simulated the event with spatial distribution varying in the intensity of precipitation from one scheme to the other . TRMM always underestimate the precipitation where its maximum was 35 mm over the Red sea while maximum precipitation from all the schemes reach up to 85 mm over different spatial domains depends on the selected used microphysics scheme. all the model schemes produced high precipitation over both Elba National park and Wadi Elgemal National park while TRMM also underestimated the both areas. Average Precipitation over Ras gab from TRMM was 7.187 mm. The closest scheme to TRMM was Morrison double moment, overestimating precipitation by 8.5 mm while the farthest scheme was Kessler by almost 14.6 mm. Average Precipitation over Wadi Elgemal from TRMM was 12.716 mm. The Lin scheme was the closest to TRMM over wadi Elgemal by overestimating the precipitation by 14.2 mm while the fahrest scheme was Eta scheme by 24.1 mm. The precipitation over Elba National park was 4.65. The closest scheme was Thompson which estimate the precipitation by 12.9. ETA scheme was the farhest one which evaluates it by 15.39 mm as shown in figure (4)

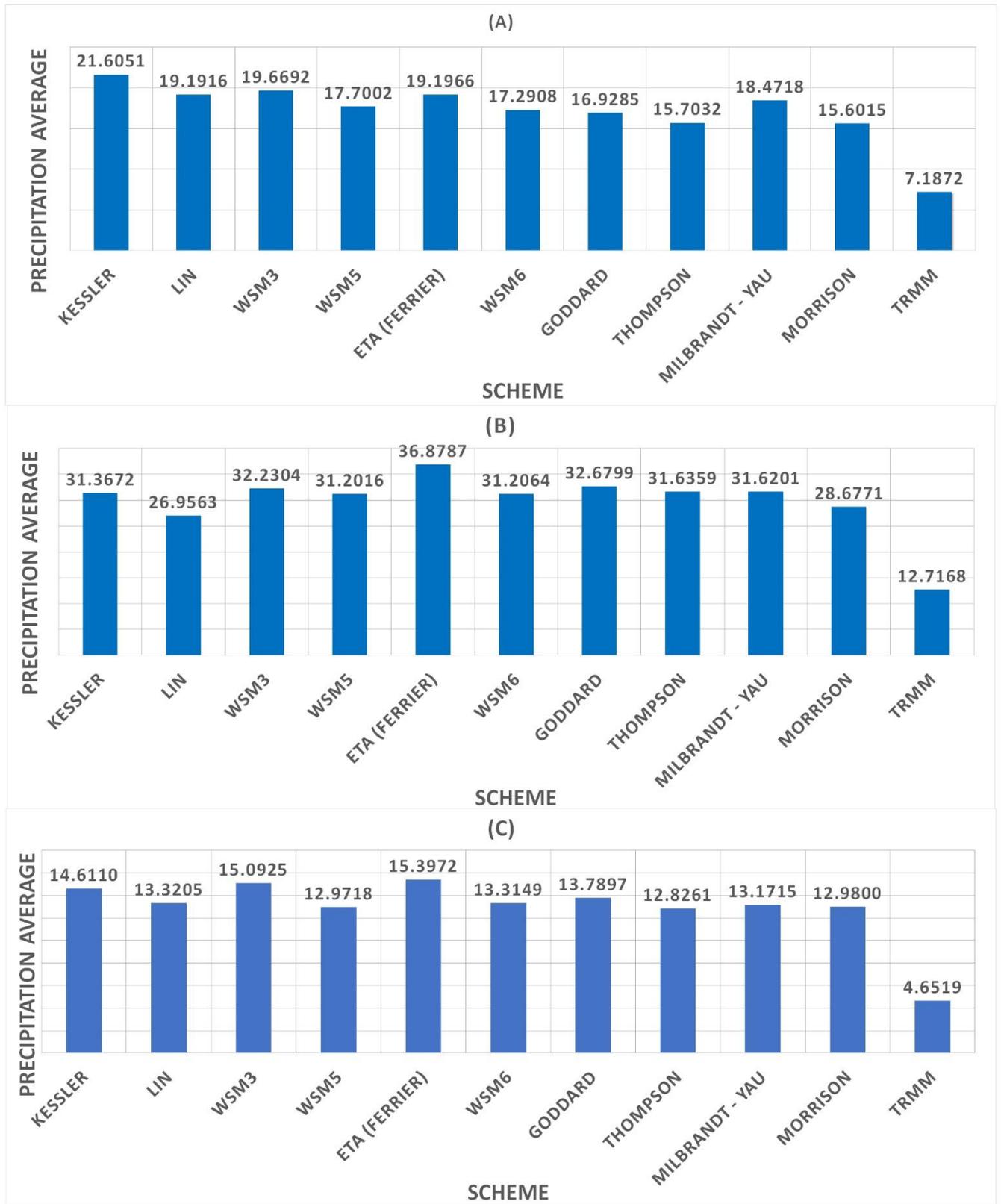

Figure (4) Average precipitation over three selected domains (A) Ras Gharib (B) Wadi Elgemal (C) Elba National park from MODIS and the ten Microphysics schemes

The spatial distribution for temperature, Pressure contours and wind speed from all ten schemes are presented also MODIS land surface temperature is shown in figure (5) at hour 12 pm. As shown that there is no values from MODIS over the upper region from the map. In terms of spatial distribution The same three selected areas for validating the model were chosen. The average Temperature over Ras Gharib was around 42.04 °C. all the schemes underestimate the average Temperature over Ras Gharib domain by almost 11°C. The closest one to MODIS was MILBRANDT-YAU Double Moment scheme which estimates the average temperature by 32.55°C. The average temperature over Wadi Elgemal domain was 40.44 °C. All the schemes are underestimate MODIS by around 13°C. Over Elba National park's domain, MODIS data estimated the temperature to be 36.88°C. while all the schemes overestimate the Temperature over this domain by percentage around 6 % as shown in figure (6).

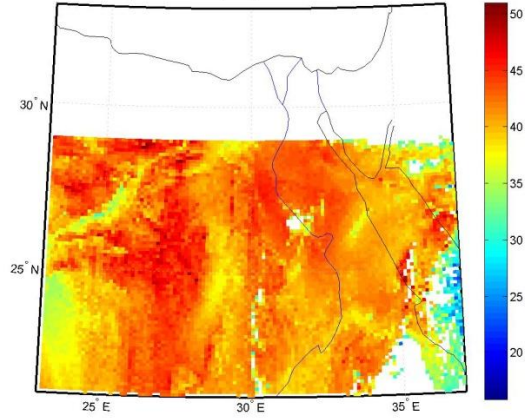
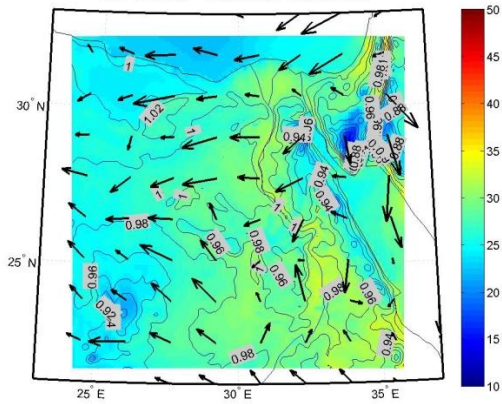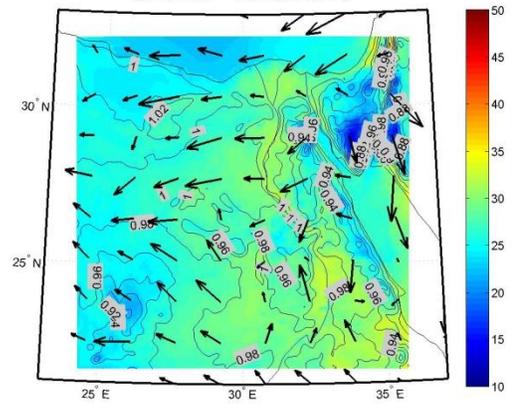
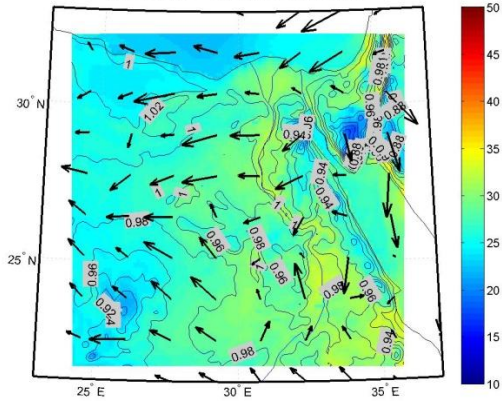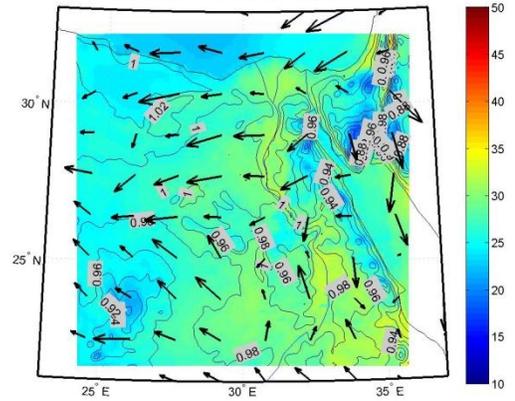

Figure (5) Spatial distribution for Temperature from MODIS against spatial distribution for Temperature, pressure, and wind directions. Overall, MODIS overestimates the Temperature over most of Egypt areas.

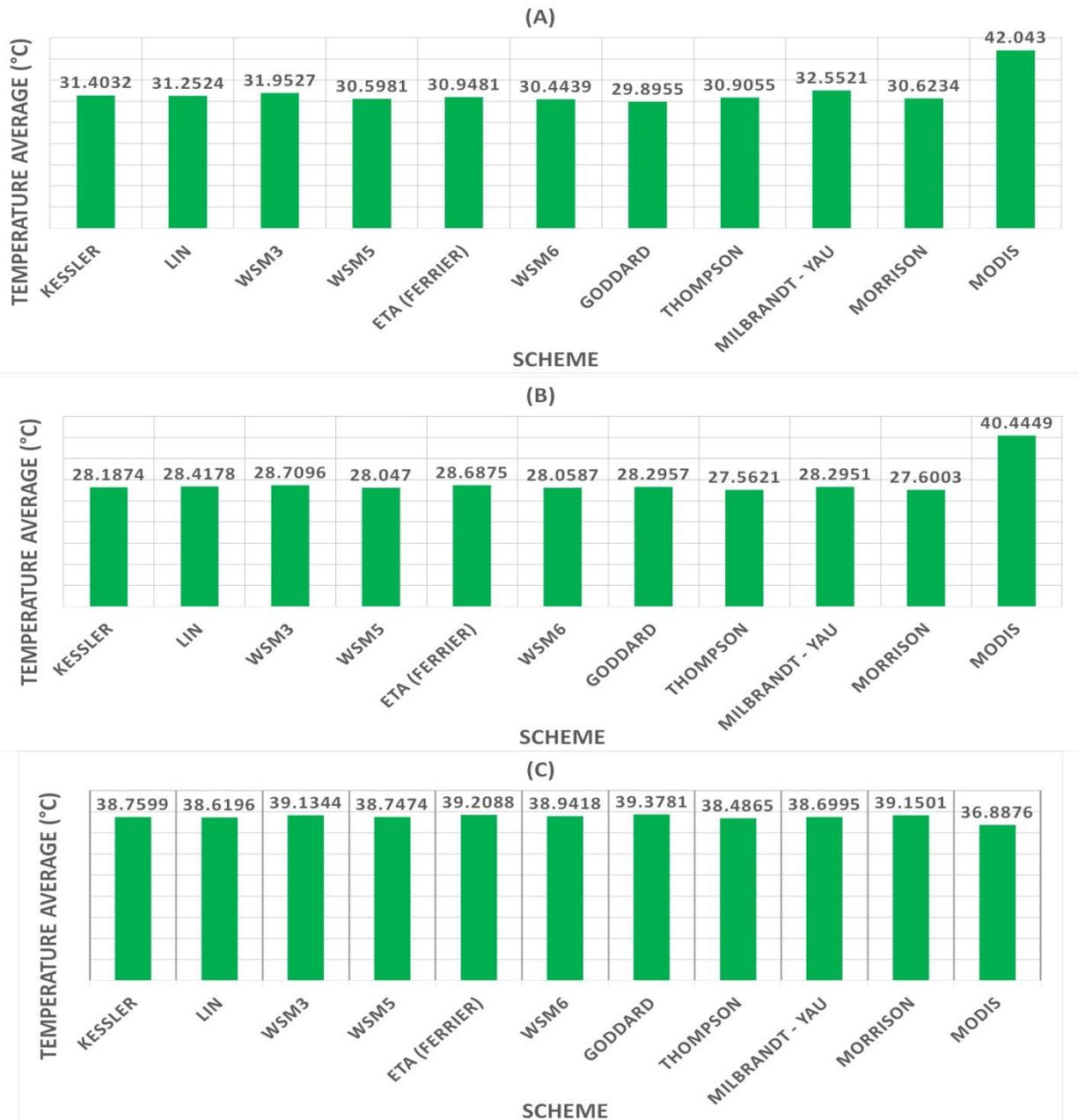

Figure (6) Average Precipitation over three selected domains (A) Ras Gharib (B) Wadi Elgemal (C) Elba National park from MODIS and the ten Microphysics schemes

The model also successfully modeled the cloud pattern as shown in figure (7) using Thompson scheme. This pattern is a function of the model height layer and the height layer is divided into three domains. Low cloud starts from the surface up to 0.66 eta level. Mid cloud from 0.66 to 0.33 eta level. high cloud from 0.33 to 0 eta level. Clouds exist at all the height layers over Sinai, while there are generally clouds at low level over the coast of the red sea and on different levels at different spatial areas.

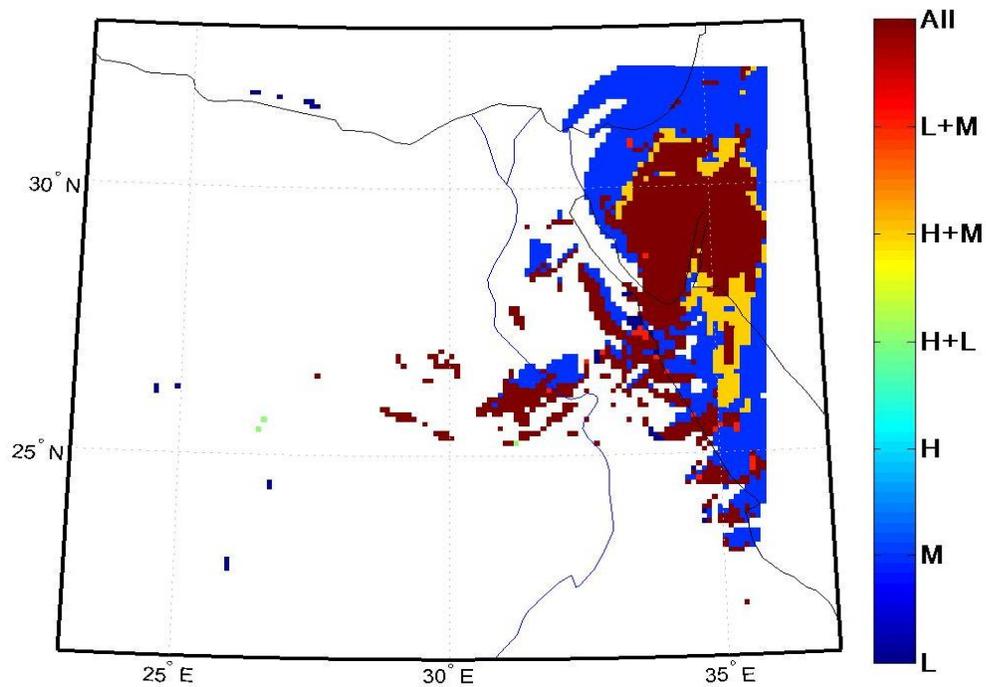

Figure (7) : Cloud pattern over Egypt using Tompshon scheme

## 5-Conclusion and future work

The WRF model was used to simulate the rainy severe event. The impact of several microphysics schemes on precipitation simulation was investigated. Performance of the WRF model was compared to satellite data (MODIS) in terms of land surface temperature and TRMM data in terms of precipitation. Three areas within the domains were selected to validate the model. WRF was able to simulate and capture the spatial pattern for precipitation, when compared to data obtained from TRMM , it was found that the model always overestimates the precipitation values. Changing the microphysics has direct impact on the spatial pattern for precipitation and on the location with highest precipitation. Not the same microphysics scheme gives the nearest values to TRMM over the three domains. For the temperature, the model produces almost the same spatial distribution for the temperature. MODIS always overestimated the temperature in all the spatial maps except over Elba National park's domain. Future works should include the testing sensitivity of the model to changes in the preferences of the cloud scheme, also the performance of the model using data assimilation techniques should be investigated.